%% file: main.tex
\documentclass[11pt]{article}
\input{preamble.tex}

\title{Committor guided estimates of molecular transition rates}
\author{Andrew R. Mitchell and Grant M. Rotskoff}
\date{\today}

\graphicspath{{figures/}}

\usepackage[sort&compress, numbers]{natbib}
\begin{document}

\maketitle

\begin{abstract}
    The probability that a configuration of a physical system reacts, or transitions from one metastable state to another, is quantified by the committor function.
    This function contains richly detailed mechanistic information about transition pathways, but a full parameterization of the committor requires building representing a high-dimensional function, a generically challenging task.
    Recent efforts to leverage neural networks as a means to solve high-dimensional partial differential equations, often called ``physics-informed'' machine learning, have brought the committor into computational reach.
    Here, we build on the semigroup approach to learning the committor and assess its utility for predicting dynamical quantities such as transition rates. 
    We show that a careful reframing of the objective function and improved adaptive sampling strategies provide highly accurate representations of the committor. 
    Furthermore, by directly applying the Hill relation, we show that these committors provide accurate transition rates for molecular system. 
\end{abstract}

\section{Introduction}

A statistically representative collection of transition pathways in a condensed phase molecular system can yield profound physical insight about a complex reaction~\cite{bolhuis_transition_2000}.
Experimentally characterizing reaction pathways with atomistic resolution in biophysical systems, however, is exceedingly difficult due to the disparate time scales and collective molecular fluctuations at play. 
Molecular dynamics simulation provides a toolkit that, in principle, provides a window into these dynamical events, but the temporal and spatial scales accessible with atomic resolution molecular dynamics lead to a complementary set of challenges. 
The kinetics of many molecular processes of interest are dominated by rare events inaccessible to current computational methods, and dimensionality reduction and approximations must thus be invoked~\cite{hall_practical_2022}.

Contemporary enhanced sampling methodology is sophisticated and diverse, but most modern methods still rely on low dimensional reaction coordinates~\cite{mehdi_enhanced_2024} which can limit the accuracy of resulting free energy surfaces.
Alternative approaches based on estimating high-dimensional committor~\cite{li_computing_2019,khoo_solving_2019, rotskoff_active_2021, li_semigroup_2022, strahan_predicting_2023, strahan_inexact_2023,jung_machine-guided_2023, evans_computing_2023,liang_probing_2023} functions have recently gained traction due to the efficacy of neural networks as representations of high-dimensional functions.
In large part, efforts to produce a representation of the committor have used data collected from enhanced sampling simulations, e.g., metadynamics~\cite{barducci_metadynamics_2011} that incorporate a low-dimensional reaction coordinate, or have leveraged data from transition path sampling to optimize a representation of the committor~\cite{jung_machine-guided_2023}, which is itself nontrivial to collect. 

Efficient data acquisition that is unbiased by a choice of reaction coordinate remains a significant bottleneck for these calculations, and our previous work~\cite{rotskoff_active_2022} incorporates adaptive importance sampling, which guides sampling using the online approximation of the committor function. 
The variational objective function in that work and others~\cite{li_computing_2019,khoo_solving_2019, rotskoff_active_2022}, however, requires diffusive, overdamped Langevin dynamics, which renders it inapplicable to most condensed phase molecular dynamics simulations, as underdamped dynamics is used to improve numerical stability.
An alternative that avoids this constraint appears in \citet{li_semigroup_2022}, but still requires the Dirichlet form and Boltzmann sampling. 
Similarly,~\citet{chen_discovering_2023} uses a semigroup formulation for the committor, representing the solution as the minimizer of a Boltzmann weighted mean-squared loss.
These works formulate the committor using the Markov semigroup for the dynamics after some lag time $\tau$, which ensures Markovianity and yields a stable objective.
Using a stopped transition operator~\cite{strahan_longtimescale_2021}, \citet{strahan_predicting_2023} showed that the committor can be formulated as the solution of a Feynman-Kac equation which uses a mean-squared error but, crucially, does not require Boltzmann reweighting, tremendously simplifying evaluation of the objective. 
Still, the utility of representations trained with these approaches has not yet been thoroughly examined, especially in the setting of molecular simulations.  

Here, we characterize the use of self-consistent committor optimization with active importance sampling and assess whether the representation can be made sufficiently accurate to compute transition rates. 
We characterize the performance in two minimal benchmark systems to analyze the effect of the assumptions we make, while also having a clear ground truth comparison. 
We obtain highly accurate rate estimates, typically an exceedingly challenging task for molecular simulations~\cite{palacio-rodriguez_transition_2022,lazzeri_molecular_2023}.
Because our approach does not postulate a low dimensional reaction coordinate, it has the ability to simultaneously represent multiple reaction channels. 

We use the Hill relation~\cite{baudel_hill_2022} to estimate transition rates, consistent with standard practice in algorithms like adaptive multi-level splitting. 
However, we find that rate estimates are extremely sensitive to imprecision in the committor near each metastable basin, because the inverse of the committor appears in the estimator.
To overcome this, we develop and employ a biased estimator for the committor; interestingly, as we demonstrate both analytically and numerically, the bias in this estimator can be controlled with modest sample sizes.
While the Hill relation applies most simply to the setting of single barrier crossing kinetics, we also test our approach on an example that features significant intermediate metastability.
Rate estimates are more computationally demanding in this setting but still accurate, and the additional information provided by the committor produces useful insight into the reaction pathways.

\section{Theory}

We consider a system with position and momenta coordinates $(\xb,\pb) \in \Omega \subset \RR^d \times \RR^d$ evolving under a stochastic dynamics in which the stationary probability measure of the configurational degrees of freedom is given by the Gibbs-Boltzmann probability measure,
\begin{equation}
    \mu(d\xb) = Z^{-1}(\beta) e^{-\beta U(\xb)} \d\xb
\end{equation}
where $\beta = 1/(k_{\rm B} T)$ and $U:\RR^d \to \RR$ is the potential energy and $Z=\int_{\Omega} e^{-\beta (U(\xb) + \tfrac12 \pb^T \mathsf{M}^{-1} \pb)} \d\xb \d\pb$ is the partition function and $\mathsf{M}$ is the mass matrix.
For molecular systems, we typically work with underdamped Langevin dynamics,
\begin{equation}
\label{eq:uld}
    \begin{cases}
        \d \xb_t = \pb_t \d t \\
        \d \pb_t = -\nabla U(\xb_t) \d t - \gamma \pb_t + \sqrt{2\gamma \beta^{-1}} \d \Wb_t
    \end{cases}
\end{equation}
where the initial positions are sampled from some initial distribution $\xb_0 \sim \rho_0$ and the momenta is selected from a Maxwell-Boltzmann distribution $\pb_0\sim \rho_{\rm MB}(\beta).$

We now develop the variational formulation of the partial differential equation defining the probability of a reactive event. 
Throughout, we focus on a transition from one metastable state $A \subset \Omega$ to a distinct state $B \subset \Omega\setminus A$.
We assume $A$ and $B$ are open, bounded subsets of $\Omega$. 
The transition rate between $A$ and $B$ can be expressed in terms of hitting times
\begin{equation}
    \tau_S(\xb_0) = \inf_{t} \{ \xb_t^{\xb_0} \in S \}
\end{equation}
where $S\subset \Omega$ and $\xb_t^{\xb}$ is the solution to initial condition $\xb_0.$
For a given configuration, the probability of reacting $A\to B$ is quantified through the \emph{forward committor function}~\cite{vanden-eijnden_transition_2005},
\begin{equation}
    q^+(\xb) = \mathbb{P} \bigl[ \tau_B(\xb) < \tau_A(\xb) \bigr].
    \label{eq:committor}
\end{equation}
where the path measure $\mathbb{P}$ is determined by the choice of stochastic dynamics. 
The backward committor function is defined analogously; when the dynamics is reversible, $q^+ = 1 - q^-$ and henceforth we denote $q:=q^+$ because we only consider equilibrium transitions in this work. 
The phase space committor, $\bar q$, solves a high-dimensional Dirichlet problem,
\begin{equation}
    \begin{cases}
        (\mathcal{L} \bar q)(\xb, \pb) = 0 \quad \xb\in \Omega \setminus (A\cup B), \\
        \bar q(\xb, \pb) = 0 \quad \xb \in A, \\
        \bar q(\xb, \pb) = 1 \quad \xb \in B,
    \end{cases}
    \label{eq:infbke}
\end{equation}
with the infinitesimal generator 
\begin{equation}
\label{eq:gen}
    \mathcal{L} = \pb \cdot \nabla_{\qb} - \nabla_{\xb} \cdot U(\xb) \nabla_{\pb}  + \gamma \left( -\pb \nabla_{\pb} + \beta^{-1} \Delta_{\pb} \right).
\end{equation}
The generator \eqref{eq:gen} is Markovian and hence provides enabling the representation for the backward Kolmogorov equation \eqref{eq:infbke}~\cite{pavliotis_stochastic_2014}.
We have shown that even high-dimensional committor functions can be determined using active importance sampling and a neural network ansatz~\cite{rotskoff_active_2021} using a PINN loss~\cite{karniadakis_physics-informed_2021}, but only in the case of overdamped dynamics.
For molecular systems, this equation could be solved directly with the infinitesimal generator for underdamped Langevin by extending the committor to the full phase space coordinates.

Generally, we want to analyze the commitment probabilities in state space, rather than the full phase space, which motivates a distinct approach that we outline now. 
If we introduce a sufficiently large lag time $\tau$, we can define an associated Markovian propagator in the state space
\begin{equation}
    \mathcal{P}_\tau f(\xb) = \EE[ f(\Xb_\tau^{\xb})],
\end{equation}
where $\mathcal{P}_\tau$ is the Markov semigroup operator associated with the dynamics and $f:\Omega\to \RR$ is some observable.
In principle, a posteriori checks to ensure that the propagator is Markovian on the timescale $\tau$ could be carried out if there is ambiguity. 
In the results shown below, we evaluate the accuracy of the committor without directly checking the Markovianity of the propagator on the timescale $\tau.$

Using the Markovian semigroup~\cite{li_semigroup_2022} provides an alternative to~\eqref{eq:infbke} and allows a distinct expression for the committor, 
\begin{equation}
    q(\xb) = \EE [ q(\Xb_\tau^{\xb}) ] \equiv \mathcal{P}_\tau q(\xb), \quad x\in \Omega \setminus (A\cup B),
    \label{eq:sceq}
\end{equation}
where the expectation is computed over dynamical trajectories of duration $\tau$.
This equation quantifies the intuitive relation that the average value of the committor for a collection of trajectories $\{ \xb_\tau^{\xb_0} \}$ evolving independently from the initial condition $\xb_0$ must equal the committor of $\xb_0$.
The relation is clearest in the limit $\tau \gg \tau_{B}(\xb_0)$, because in this case the right-hand side of~\eqref{eq:sceq} is directly an empirical estimate of the fraction of trajectories that first hit $B.$
The expression \eqref{eq:sceq} is a particular instantiation of a Feynman-Kac equation, and the general strategy for solving equations of this type with a neural network ansatz was first developed in Ref.~\cite{strahan_predicting_2023}. 
When the dynamics is described by a time-homogeneous Markov process, the transition semigroup can be written in terms of an infinitesimal generator, $\mathcal{L}$ where $\mathcal{P}_{\tau} = e^{\tau \mathcal{L}}$.
In the limit $\tau \downarrow 0$, this recovers~\eqref{eq:infbke}.

\begin{figure}
    \centering
    \includegraphics[width=\linewidth]{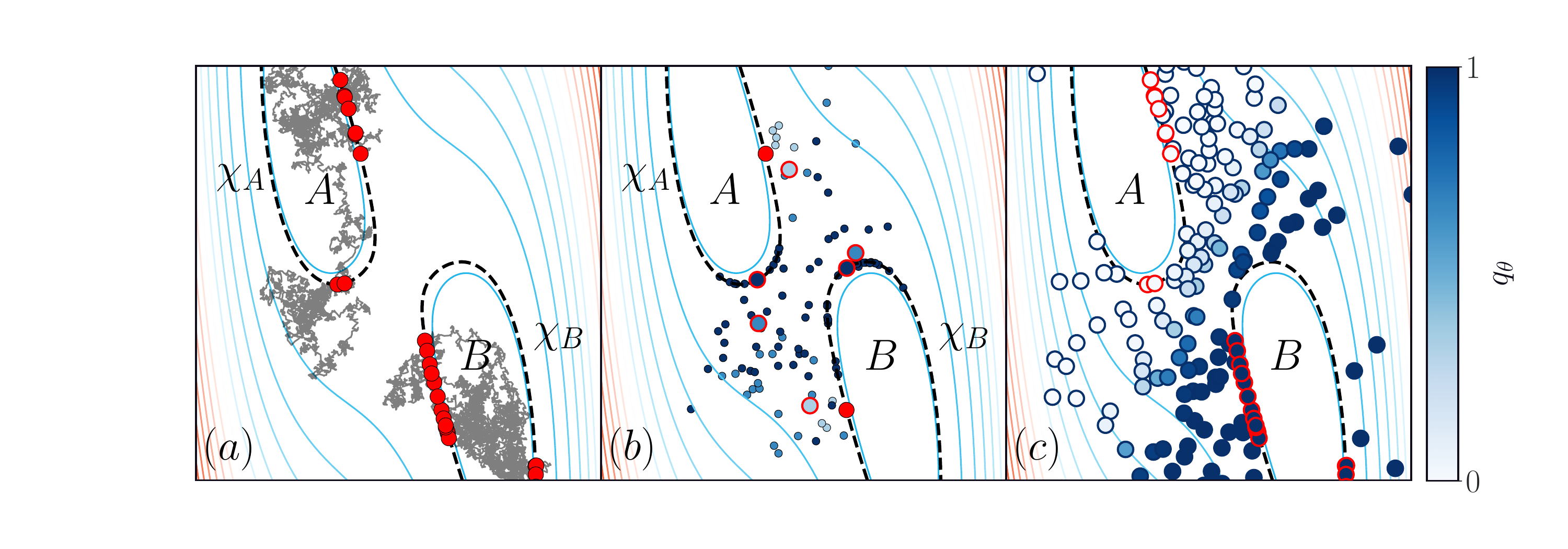}
    \caption{(a) Equilibrium simulations (black lines) are initiated in each basin, and configurations that exit the boundaries $\chi_A$ and $\chi_b$ (red dots) are collected, along with the mean time between exits. (b) From each exit configuration, swarms of trajectories are launched and their endpoints (black borders) are used to train the committor. The sampling points from which the next swarm will be generated (red borders) are chosen from previous swarm endpoints based on their estimated committor values. Once an unbiased transition path is sampled, the algorithm repeats starting with a new exit configuration. (c) After many iterations, this algorithm collects many samples in the transition state region (blue borders) and an accurate global committor is learned; the committor is then evaluated at the exit configurations from (a) (red borders) and combined with the mean exit time to compute a rate estimate via \eqref{eq:hill}.}
    \label{fig:schematic}
\end{figure}

While the committor is often heralded as the ``ideal reaction coordinate'', it also enables efficient estimates of transition rates~\cite{lopes_analysis_2019}.
To estimate rates, we employ the Hill relation~\cite{kramers_brownian_1940, baudel_hill_2022}, which decomposes the transition rate into an outward flux from basin $A$, denoted $j_{A^-}$ and reactive probabilities at the boundary, 
\begin{equation}
    k_{AB} = j_{A^-} \int_{\chi_A} q(\xb) \d \xb.
    \label{eq:hill}
\end{equation}
This formalism is closely related to the widely used forward flux sampling algorithm~\cite{allen_forward_2009}, transition interface sampling~\cite{vanerp_elaborating_2005, hall_practical_2022}, and adaptive multi-level splitting~\cite{cerou_adaptive_2007}.
In each of these approaches, the integral on the right-hand side of~\eqref{eq:hill} is estimated by reweighting transitions between a sequence of interfaces.
With a direct parameterization of the committor, however, this estimate is considerably simpler.
The general computational approach is illustrated in Fig.~\ref{fig:schematic} and discussed in detail in Sec.~\ref{sec:computational}.

In this work, we optimize a variational objective to solve~\eqref{eq:sceq} using a neural network ansatz for the committor function $q$ in the configuration space for biomolecular systems, allowing rate estimates using~\eqref{eq:hill}.
To do so, we must create an objective function that, when minimized, satisfies~\eqref{eq:sceq}. 
In particular, we represent the committor as a parametric function, $q_{\theta}:\Omega\to [0,1]$ and solve
\begin{equation}
    q_{\star} = \min_{\theta} L(\theta).
    \label{eq:poploss}
\end{equation}
Producing an objective function for \eqref{eq:sceq} is straightforward in this case; previous works \cite{ strahan_predicting_2023, chen_discovering_2023} have used a mean-squared error.
Our approach differs in an important way, we instead use a mean-squared discrepancy on the $\log$-committor so that
\begin{equation}
    L(\theta) = \frac12 \int_{\Omega} \bigl( \log(q_{\theta}(\xb)) - \log(\mathcal{T}_\tau^{\Omega^{\rm c}} q_{\theta}(\xb)) \bigr)^2 \d \nu(\xb),
    \label{eq:objective}
\end{equation}
where the expectation is computed over initial conditions drawn from some sampling distribution $\nu$ which is not necessarily the equilibrium measure $\mu.$
In this equation, $\mathcal{T}^{\Omega^{\rm c}}_\tau$ denotes the stopped transition operator~\cite{strahan_longtimescale_2021}, which terminates when hitting $A\cup B$; 
formally, it is defined as $\EE f(\Xb_t^{\xb})$ 
where $t = \textrm{min} \left[ \tau, \tau_{A\cup B}(\xb) \right]$.
While~\eqref{eq:objective} carries a statistical bias, we show in Sec.~\ref{sec:loss}, that this bias is controlled and, in fact, as detailed below, the use of this biased loss improves our rate estimates.

\section{Logarithmic Committor Loss}
\label{sec:loss}

Accurately estimating transition rates is the central goal of our approach, which in turn requires accurate committor probabilities near the reactant state $A.$
When the timescale for a typical reaction to occur is much longer than the molecular relaxation time, typical committor values near the metastable state $A$ will concentrate near zero. 
Therefore, the logarithmic difference between $q_{\theta}(\xb)$ and $\mathcal{P}_\tau q_{\theta}(\xb)$ in the objective scales these small probabilities to prioritize this region of configuration space.  

The success of our algorithm at predicting the committor in regions close to the basins relies on the logarithmic bias in \eqref{eq:objective}, and we analyze this bias to show that it is controlled.
Using~\eqref{eq:sceq}, the action of the semigroup can be estimated by the unbiased sample mean, giving
\begin{equation}
    q(\xb) \approx \frac{1}{n} \sum_{i=1}^{n} q(\xb_\tau^{\xb, i}) \coloneq \hat{q}(\xb)
    \label{eq:emp_ck}
\end{equation}
for a collection of $i$ endpoints $\xb_\tau^{\xb, i}$ of trajectories of duration $\tau$ initialized from $\xb$. Taking the logarithm of both sides yields our estimator for the logarithm of the committor
\begin{equation}
    \log q(\xb) \approx  \log \left ( \frac{1}{n} \sum_{i=1}^{n} q(\xb_\tau^{\xb, i}) \right ) = \log \hat{q}(\xb).
    \label{eq:logq}
\end{equation}
However, this estimator is biased, as is revealed by Jensen's inequality,
\begin{equation}
        \EE [\log\hat{q}(\xb)] \leq  \log \bigl( \EE [\hat{q}(\xb)] \bigr)
        = \log q(\xb).
    \label{eq:jensen}
\end{equation}
We aim to show that this bound is tight for the committor, or, equivalently, that the ``Jensen gap'',
\begin{equation}
        \frac{\EE [\log\hat{q}(\xb)]}{\log \bigl(\EE [\hat{q}(\xb)]\bigr)} \approx 1.
    \label{eq:jensen_gap}
\end{equation}

\begin{figure}
    \centering
    \includegraphics[width=0.7\linewidth]{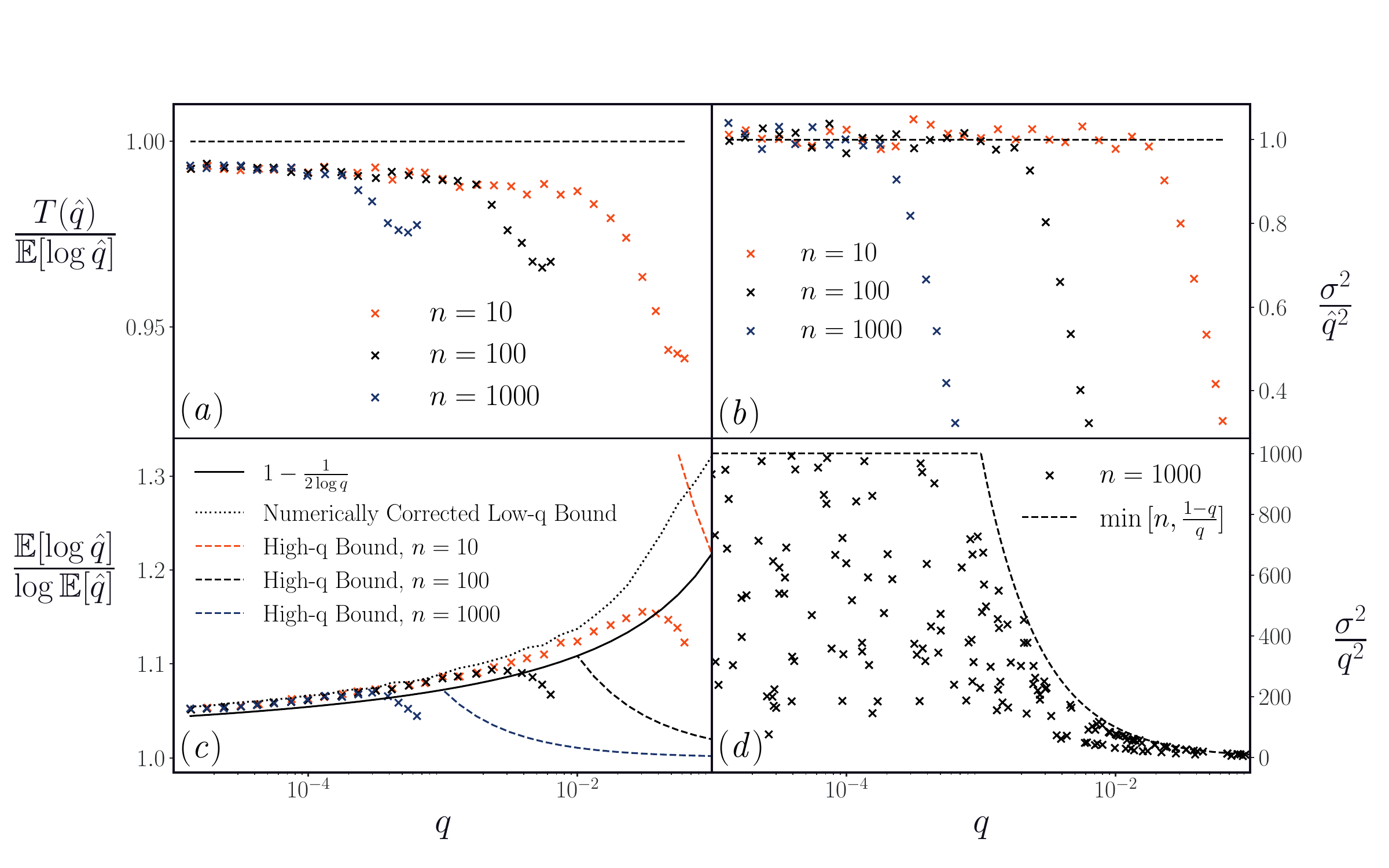}
    \caption{($a$) Relative error in the second-order Taylor approximation \eqref{eq:taylor} for the sample mean estimator $\hat{q}$ of exponential random variables (chosen to maximize the coefficient of variation at small $q$) at various means $q$ and sample sizes $n$. ($b$) Coefficient of variation for the sample mean estimator $\hat{q}$ for the same set of exponential variables. ($c$) Jensen gap for the same set of exponential variables, with upper bounds \eqref{eq:jensen_gap_3} and \eqref{eq:jensen_gap_largeq}. The variables shown overshoot the low-$q$ bound due to small underestimates in the Taylor approximation; dividing the bound by the relative error in the approximation corrects this. ($d$) Coefficient of variation for sets of trajectory endpoints generated by our algorithm in a two-channel potential. The coefficient clearly respects a large-$q$ bound.}
    \label{fig:Bias}
\end{figure}

To show this, we assume that for $n$ large enough, the Central Limit Theorem (CLT) assures that $\hat{q}(\xb)$ is normally distributed with mean $q(\xb)$ and variance $\frac{\sigma^2}{n}$, where $\sigma^2(\xb)$ is the variance of the committor evaluated at the end of a trajectory of fixed length $\tau$ initialized at $\xb$. 
In the case of the committor, which takes on bounded non-negative values, the coefficient of variation for the collection of trajectory endpoints is bounded above by the sample size \cite{katsnelson_upper_1957},
\begin{equation}
        \frac{\sigma(\xb)}{q(\xb)} \leq \sqrt{n},
    \label{eq:covbound}
\end{equation}
and, accordingly, the bound on the coefficient of variation for the estimator $\hat{q}$ is
\begin{equation}
        \frac{\sigma(\xb)}{\sqrt{n}q(\xb)}\leq 1.
    \label{eq:covqhatbound}
\end{equation}
As a result, we can expand $\log \hat{q}(\xb)$ around its mean $q(\xb)$ to obtain,
\begin{equation}
    \begin{split}
        \EE [\log \hat{q}(\xb)] & = \EE[q(\xb)] + \EE \left [ \frac{\hat{q}(\xb) - q(\xb)]}{q(\xb)} \right ] - \EE \left [ \frac{(\hat{q}(\xb) - q(\xb))^2}{2q(\xb)^2} \right ] + \dots \\
        & \approx \log q(\xb) - \frac{\sigma^2(\xb)}{2n q(\xb)^2} 
    \end{split}
    \label{eq:taylor}
\end{equation}
The Jensen gap then reduces to
\begin{equation}
        \frac{\EE [\log \hat{q}(\xb)]}{\log (\EE [\hat{q}(\xb)])} \approx \frac{\log(q(\xb)) - \frac{\sigma^2(\xb)}{2n q(\xb)^2}}{\log(q(\xb))} = 1 - \frac{\sigma^2(\xb)}{2n q(\xb)^2\log(q(\xb))},
    \label{eq:jensen_gap_2}
\end{equation}
and due to the bound on the coefficient of variation \eqref{eq:covqhatbound}, the upper bound as a function of $q(\xb)$ is
\begin{equation}
        \frac{\EE [\log(\hat{q}(\xb))]}{\log (\EE [\hat{q}(\xb)])} \leq 1 - \frac{1}{2\log(q(\xb))}.
    \label{eq:jensen_gap_3}
\end{equation}

For points near the basin, where $q(\xb)\ll 1$, the Jensen gap ensures that the bias is well controlled. It may appear from this bound that the use of the logarithmic objective is inappropriate for large values of $q(\xb)$; however, as $q(\xb)$ becomes larger, stricter bounds on the coefficient of variation \eqref{eq:covbound} become applicable. Specifically, the bounded nature of the committor restricts the endpoint variance $\sigma^2(\xb)$ to no greater than $q(\xb)(1-q(\xb))$; this bounds the coefficient of variation by $\sqrt{\frac{1-q(\xb)}{q(\xb)}}$ leading to another upper bound on the Jensen gap

\begin{equation}
        \frac{\EE [\log(\hat{q}(\xb))]}{\log (\EE [\hat{q}(\xb)])} \leq 1 - \frac{1 - q(\xb)}{2nq(\xb)\log(q(\xb))}
    \label{eq:jensen_gap_largeq}
\end{equation}
which is loose at small $q(\xb)$ but is tighter than \eqref{eq:jensen_gap_3} at larger values. For example, at the transition state, where $q(\xb) = 1 - q(\xb) = 0.5$, the upper bound becomes 
\begin{equation}
        \frac{\EE [\log(\hat{q}(\xb))]}{\log (\EE [\hat{q}(\xb)])} \leq 1 + \frac{1}{2n\log(2)},
    \label{eq:jensen_gap_ts}
\end{equation}
which is negligible for large n.

By identifying the intersection of the bounds \eqref{eq:jensen_gap_3} and \eqref{eq:jensen_gap_largeq}, an absolute upper bound independent of $q(\xb)$ can be calculated. The intersection occurs when the two upper bounds on the coefficients of variation coincide
\begin{equation}
        \sqrt{n} = \sqrt{\frac{1-q(\xb)}{q(\xb)}} \rightarrow q(\xb) = \frac{1}{n+1},
    \label{eq:equal_cov_bounds}
\end{equation}
leading to the absolute bound 
\begin{equation}
        \frac{\EE [\log(\hat{q}(\xb))]}{\log (\EE [\hat{q}(\xb)])} \leq 1 + \frac{1}{2\log(n+1)}.
    \label{eq:jensen_gap_4}
\end{equation}
Surprisingly, the bias is at a maximum at this intermediate value of $q(\xb)$, and ameliorates slowly as $q(\xb) \rightarrow 0$ and sharply as $q(\xb) \rightarrow 1$, as shown in Figure \ref{fig:Bias}.

Although the bias is bounded, there is a systematic underestimate of $q(\xb)$ present on the order of a few percent. Because we symmetrize our loss function \eqref{eq:emploss} in practice, this ultimately suggests that our learned committor $q_{\theta}(\xb)$ may be slightly more peaked around the transition state than the true committor, leading to a slight underestimation of rates in both directions.

\section{Computational Details}
\label{sec:computational}

We optimize the committor and estimate the rate ``on-the-fly''. 
The algorithm we employ combines local sampling within the reactant and product basins with importance sampling to obtain data for the committor away from the basins. 
This section systematically outlines the computational approach and details how we integrate the committor update with the rate estimates.
The algorithm we use requires many independent simulations of short, ``swarm'' trajectories~\cite{pan_finding_2008}, which are trivially parallelizable. 

First, an equilibrium simulation is run in each basin until each trajectory has exited the basin $N_{\rm exit}$ times. 
Each time the trajectory exits the basin, its configuration and the time at which the exit took place are stored. 
This information constitutes an ensemble of points sampled from the basin boundaries $\chi_A$ and $\chi_B$ and exit times that can be used to calculate the fluxes $j_{A^-}$ and $j_{B^-}$. The uncertainty in these estimates, and subsequently the uncertainty in the estimated transition rate, depends on the number of exit events observed.

The importance sampling scheme we employ selects points from one boundary ensemble and attempts to sample points along a reactive trajectory, eventually sampling points near the opposite basin. First, a boundary configuration $\xb_i$ from one basin is chosen uniformly at random, from which a swarm of $k$ independent trajectories $\xb_\tau^{\xb_i, j}$ is initiated and evolved. The sample $\xb_i$ and the endpoints of each trajectory in its associated swarm $\xb_\tau^{\xb_i, j}$ are saved and used to train the committor $q_\theta$ via $n_{\textrm{iter}}$ iterations of stochastic gradient descent on the symmetrized objective
\begin{equation}
    \begin{split}
        L_{n,k,\tau}(\theta) &= \frac{1}{2n} \sum_{i=1}^n \left( \log(q_{\theta}(\xb_i)) - \log(\frac{1}{k}\sum_{j=1}^k q_{\theta}(\xb_{\tau}^{\xb_i,j})) \right)^2 \\
        &+ \frac{1}{2n} \sum_{i=1}^n \left( \log(1 - q_{\theta}(\xb_i)) - \log(\frac{1}{k}\sum_{j=1}^k  1 - q_{\theta}(\xb_{\tau}^{\xb_i,j})) \right)^2.
        \label{eq:emploss}
    \end{split}
\end{equation}
where the conditions $q_{\theta}(\xb \in A) = 0$ and $q_{\theta}(\xb \in B) = 1$ are enforced explicitly. This estimator requires sampling $k$ trajectories from the same initial condition $\xb_i$, which is straightforward in molecular simulations. Nevertheless, this requirement can be relaxed, as pointed out in~\cite{strahan_inexact_2023}. It is also important to note that in our implementation, we detach the "target" committor value $\frac{1}{k}\sum_{j=1}^k q_{\theta}(\xb_{\tau}^{\xb_i,j}))$ from the computational graph and treat it as a nonparametric value. Thus, gradient updates are based only on the "estimate" $q_{\theta}(\xb_i)$. While in theory it is possible to re-compute $\frac{1}{k}\sum_{j=1}^k q_{\theta}(\xb_{\tau}^{\xb_i,j}))$ periodically during this training process as the committor evolves, we choose to compute it only once at the outset of each training step to ensure stability.

Our algorithm chooses the next sampling point $\xb_{i+1}$ as the swarm endpoint $\xb_\tau^{\xb_i, j}$ with the highest committor value under $q_\theta$. This algorithm resembles forward propagation strategies that underlie forward flux sampling and steered transition path sampling~\cite{hall_practical_2022, cerou_adaptive_2007, vanerp_elaborating_2005,guttenberg_steered_2012}. 
Another swarm is generated from $\xb_{i+1}$, and the committor is updated again based on a dataset now consisting of $i+1$ sampled points and their associated swarms. This process continues iteratively, with the next configuration chosen from the set of \textit{all} previously sampled trajectory endpoints, the configuration with highest committor value being chosen as the next sample. Allowing the algorithm to sample a configuration from any previous swarm affords flexibility in the construction of a transition pathway as the committor $q_\theta$ evolves and estimates of transition probabilities change.

As the committor evolves and we sample configurations with larger estimated committor values, members of a swarm should terminate in the basin opposite the basin from which the original sample $\xb_i$ was drawn. When we reach this point, the algorithm has sampled configurations in an intermediate region of state space connecting basins $A$ and $B$. We refer to this collection of samples connecting the basins as one sampling ``chain''. Next, to continue to facilitate sampling in intermediate regions of state space, a new boundary sample is randomly drawn and a new chain of samples is generated. After we acquire new samples, the committor $q_\theta$ is updated using stochastic gradient descent, evaluating the loss function \eqref{eq:emploss} on all previously sampled configurations. 
Crucially, when choosing the next configuration in a given chain, our algorithm only considers swarm endpoints generated from the current chain, and not previous chains. This prevents sampling from concentrating near the opposite basin and encourages exploration of other potential reaction pathways.

In practice, we find that it is helpful to simultaneously sample points originating from both basins $A$ and $B$. 
This allows the committor to learn the boundary conditions $q_\theta(\xb \in A) = 0$ and  $q_\theta(\xb \in B) = 1$ at the outset of sampling, as we do not enforce these boundary conditions explicitly in the objective as in other approaches \cite{rotskoff_active_2022, li_computing_2019, li_semigroup_2022}. Thus, in one iteration of our algorithm, we collect two sample points, each associated with a chain originating from each basin.

At any point in this process, rates $k_{AB}$ and $k_{BA}$ can be estimated in an online manner using \eqref{eq:hill}; the flux is estimated from the initial equilibrium simulation, and mean committor value on the boundary basin can be estimated by evaluating $q_\theta$ on the associated boundary ensemble. The convergence of the rate estimates for both transitions is a useful proxy for the convergence of $q_\theta$ itself, and we demonstrate below that rate estimates generated by this algorithm converge stably in applications to biomolecular systems.

\section{Results and Discussion}
\label{sec:results}

We applied our methodology to two benchmark systems to assess the assumptions of the algorithm.
We chose these relatively simple examples because ground truth rates from unbiased simulations are straightforward to estimate. 
We represent the committor with a minimalistic neural network ansatz because we wanted to focus attention on the sampling procedure. 
We embed the coordinates using a flattened vector of all pairwise distances: for alanine dipeptide, we used only heavy atoms, leading to an input dimension $d=45$; for Aib9, we used all atoms alongside its $18$ dihedral angles leading to an input dimension $d=8274.$
We used a multilayer perceptron with three hidden layers and 100 neurons per layer using a leaky ReLU nonlinear activation function to represent the committor $q$.
Because the output represents a probability, we pass the final output through a sigmoid function to ensure that the range is $(0,1).$

\subsection{Case study: Overdamped versus Underdamped Dynamics for Alanine Dipeptide}

First, to assess the effect of employing underdamped dynamics using the semigroup objective~\eqref{eq:sceq}, we directly compare rate calculations for a peptide fragment in the underdamped and overdamped case. 
We simulated alanine dipeptide with the Charmm27 force field~\cite{vanommeslaeghe_charmm_2010} in vacuum at 300K, which is a widely used and straightforward benchmark system for enhanced sampling methods. 
Though it is not a particularly challenging benchmark, it offers a useful comparison because it is tractable even when using the 0.01 femtosecond time step required for numerical stability with an overdamped integrator. 

\begin{figure}
    \centering
    \includegraphics[width=0.8\linewidth]{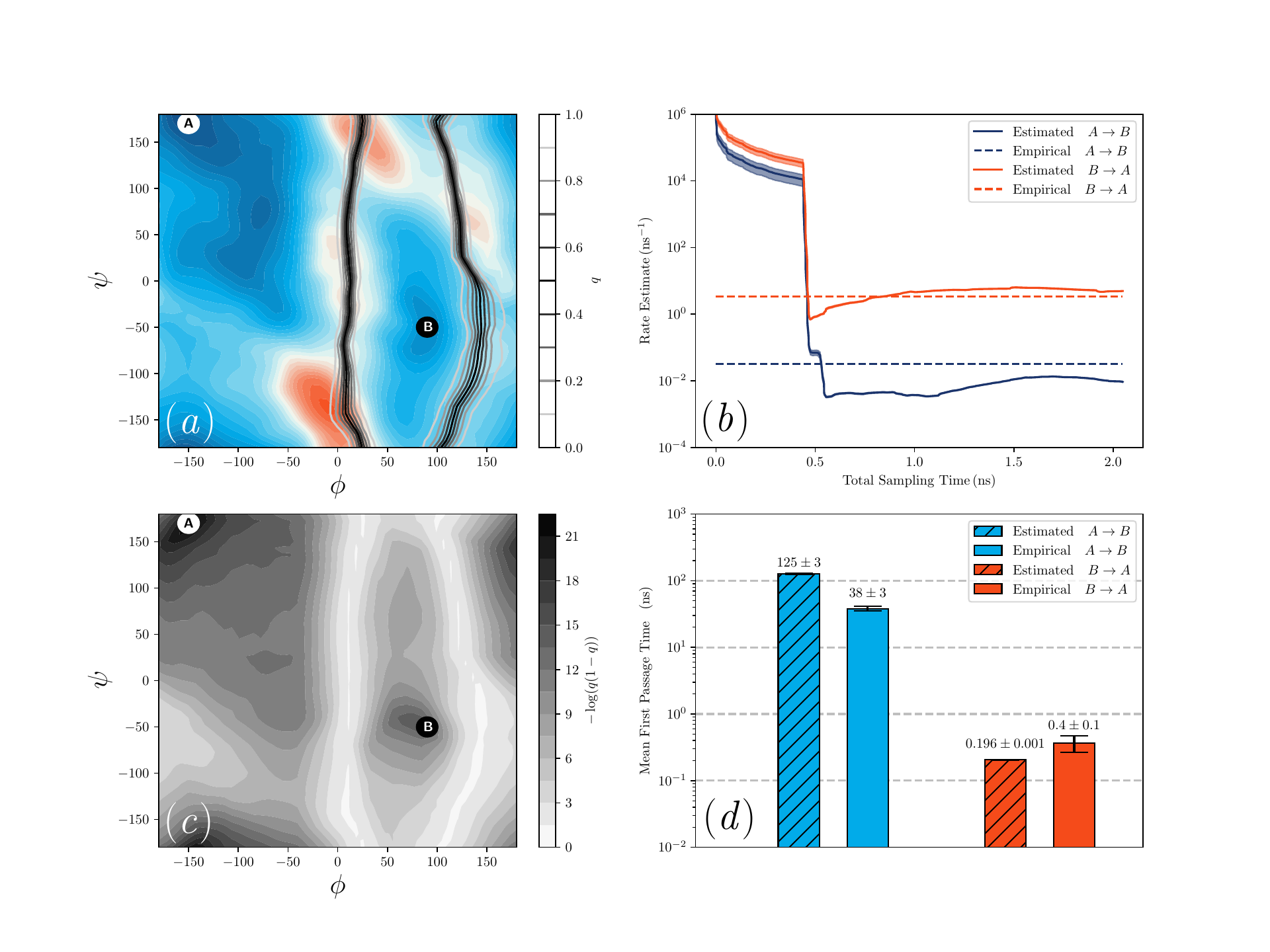}
    \caption{Results on overdamped alanine dipeptide. (a) Committor isosurfaces on the pontential of mean force in $[\phi, \psi]$ - dihedral space. (b) Running rate estimates for both directions of the $C_7^{\rm ax} \leftrightarrow C_7^{\rm eq}$, represented as the average rate estimate from the previous 500 steps of the algorithm. Rate estimates from a long equilibrium simulation are shown for comparison. (c) Relative transition probability $q(1-q)$ on a logarithmic scale in $[\phi, \psi]$ - dihedral space. (d) Final estimates of mean first passage times compared with estimates from a long equilibrium trajectory.}
    \label{fig:adp_rate_overdamped}
\end{figure}

In Fig.~\ref{fig:adp_rate_overdamped} ($a$), we show the high-dimensional committor projected onto the backbone dihedrals.
The potential of mean force is shown as filled contours, which we obtained using umbrella sampling.
The transition isosurface where $q=\tfrac12$ sits at the transition states for the $C_7^{\rm ax} \leftrightarrow C_7^{\rm eq}$ isomerization, consistent with expectations.
Interestingly, it captures all reaction pathways accurately.
We also plot the quantity $q(1-q)$ in Fig.~\ref{fig:adp_rate_overdamped} ($b$), which obtains its maximal value when $q=\tfrac12$ and symmetrizes the reactive probability flow lines.
By taking the negative logarithm of this quantity, we see that the variation in probability near basins is well resolved. 

Using the online estimate of the committor together with the reactive flux out of the basins $A$ and $B$, we estimate transition rates.
Fig.~\ref{fig:adp_rate_overdamped} ($c$) shows a rolling average of the previous 500 rate estimates as a function of total sampling time. 
At early times, the representation of the committor function is imprecise, causing substantial error in the rate. 
As the representation of the committor improves, the rate stabilizes within the correct order of magnitude. 
Note that the mean first passage time for the transition is ~38 ns and with 2 ns of total sampling time, we obtain accurate rate estimates. 

\begin{figure}
    \centering
    \includegraphics[width=0.8\linewidth]{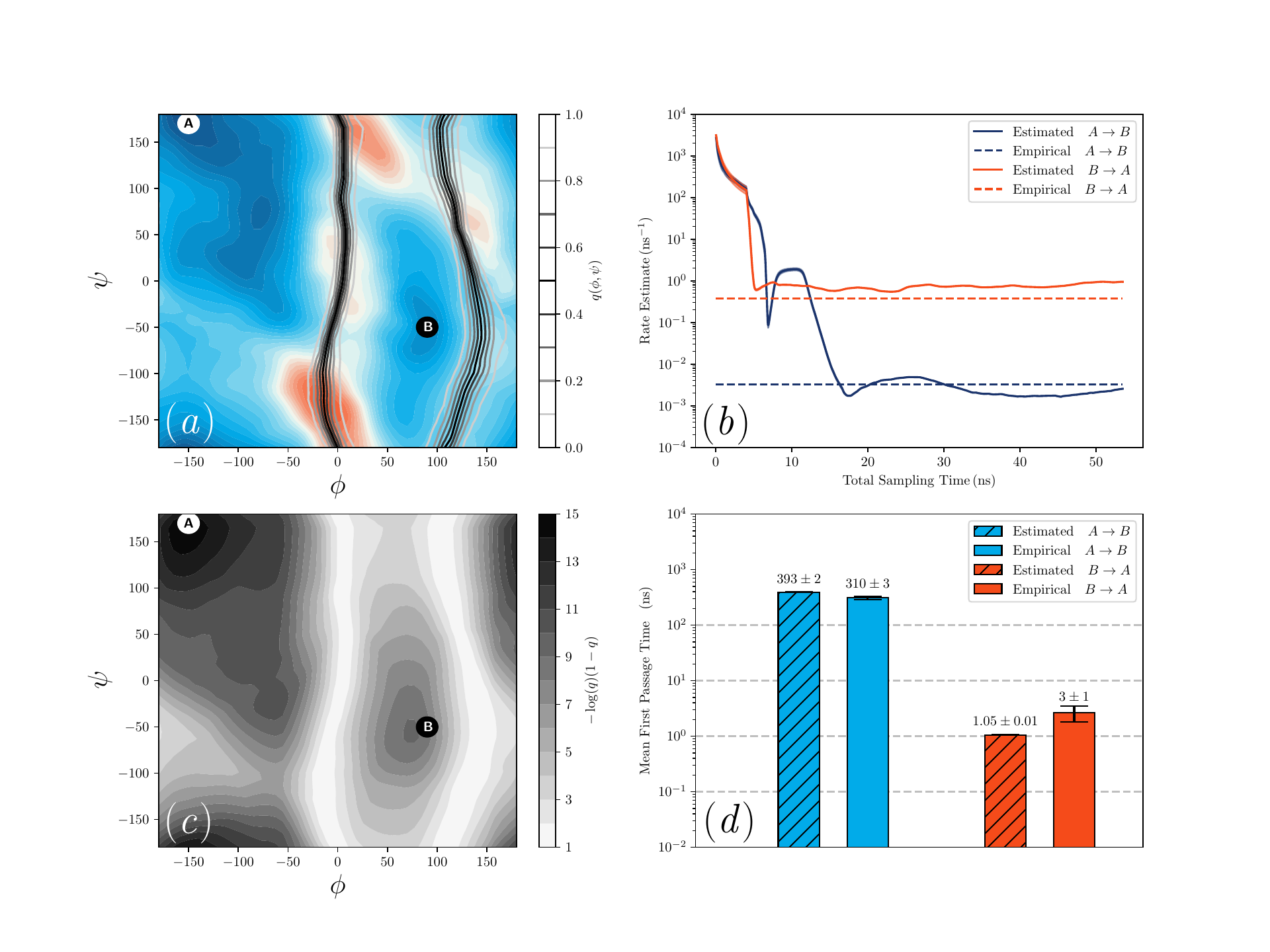}
    \caption{Results on underdamped alanine dipeptide. (a) Committor isosurfaces on the pontential of mean force in $[\phi, \psi]$ - dihedral space. (b) Running rate estimates for both directions of the $C_7^{\rm ax} \leftrightarrow C_7^{\rm eq}$, represented as the average rate estimate from the previous 200 steps of the algorithm. Rate estimates from a long equilibrium simulation are shown for comparison. (c) Relative transition probability $q(1-q)$ on a logarithmic scale in $[\phi, \psi]$ - dihedral space. (d) Final estimates of mean first passage times compared with estimates from a long equilibrium trajectory.}
    \label{fig:adp_rate_underdamped}
\end{figure}

The underdamped case, for which it is difficult to guarantee the assumption of Markovianity a priori, shows remarkably similar performance. In the case of underdamped dynamics, we initialize each trajectory in a swarm with a velocity randomly sampled from the Maxwell-Botlzmann distribution, thus effectively integrating out the velocity portion of phase space, and we continue to condition our estimate of the committor on the configuration of the molecule only.
As shown in Fig.~\ref{fig:adp_rate_underdamped} ($a$) and ($b$), we obtain committor isosurfaces that sit near the saddles of the free energy surface plotted in dihedral coordinates. 
The high energy saddle near $(100^{\rm o},100^{\rm o})$ is not well resolved, but this pathway contributes negligibly to the overall reactive flux. 
The moving average of the online rate estimate again starts with a significant overestimate, but as the committor representation is trained, the predicted rate approaches a value close to the empirically measured rate for both the $A\to B$ and $B\to A$ transition. 
As before, the total sampling time required to predict both transition rates accurately is a small fraction of the mean first passage time. 
Of course, the simulations run for our rate estimator are also trivially parallelizable. 

\subsection{Estimating transition rates with metastable intermediates: Aib9 Isomerization}

We also examined the performance of our rate estimation algorithm on an example that features challenges more characteristic of complex biophysical dynamics. 
While the system is still small, the synthetic peptide Aib9 features intermediate metastability~\cite{lorpaiboon_accurate_2024}, which complicates the computation of a rate. 
As a result, in this case, we do not fix a constant $\tau$ and instead incorporate a random stopping time for the swarms, which is contingent upon basin membership for at least one configuration.
We simulated this system using the Amber15-IPQ force field~\cite{bogetti_twist_2020} in vacuum at 500K.

\begin{figure}
    \centering
    \includegraphics[width=0.8\linewidth]{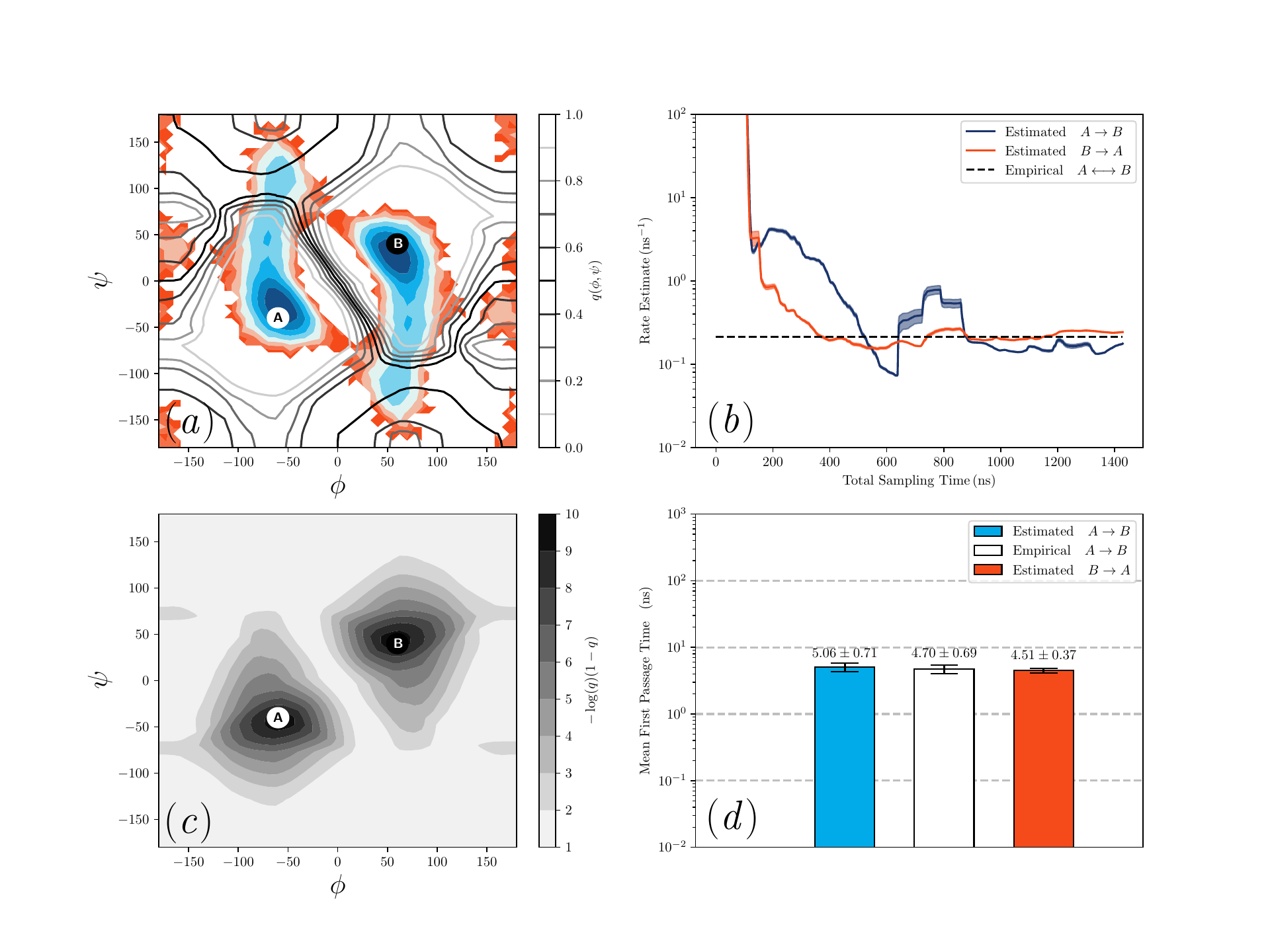}
    \caption{Results on the AIB9 system. (a) Committor isosurfaces on the pontential of mean force in the $[\phi, \psi]$ - dihedral space of the middle residue. (b) Running rate estimates for both directions of helix isomerization, represented as the average rate estimate from the previous 200 steps of the algorithm. The rate estimate from a long equilibrium simulation are shown for comparison. (c) Relative transition probability $q(1-q)$ on a logarithmic scale in the $[\phi, \psi]$ - dihedral space of the middle residue. (d) Final estimates of the mean first passage time compared with estimates from a long equilibrium trajectory.}
    \label{fig:aib9}
\end{figure}

In Fig.~\ref{fig:aib9} ($a$) and ($b$) we show the estimated high-dimensional committor projected onto the backbone dihedral angles of the fifth amino acid in the nine amino acid chain.
The potential of mean force for this coordinate is shown as filled contours for reference, which we obtained from a long unbiased simulation.
Because of the metastable intermediates, the total sampling time needed to estimate the transition rate exceeds the mean first passage time by an order of magnitude.
This occurs because we need values of the lag time $\tau$ that exceed the residence time in the intermediate metastable states.
While a naive implementation of our algorithm is unlikely to be computationally efficient in this case, the rate estimates we make, shown in Fig.~\ref{fig:aib9} ($c$) and ($d$), are within error of the empirical estimate. 

\begin{figure}
    \centering
    \includegraphics[width=0.7\linewidth]{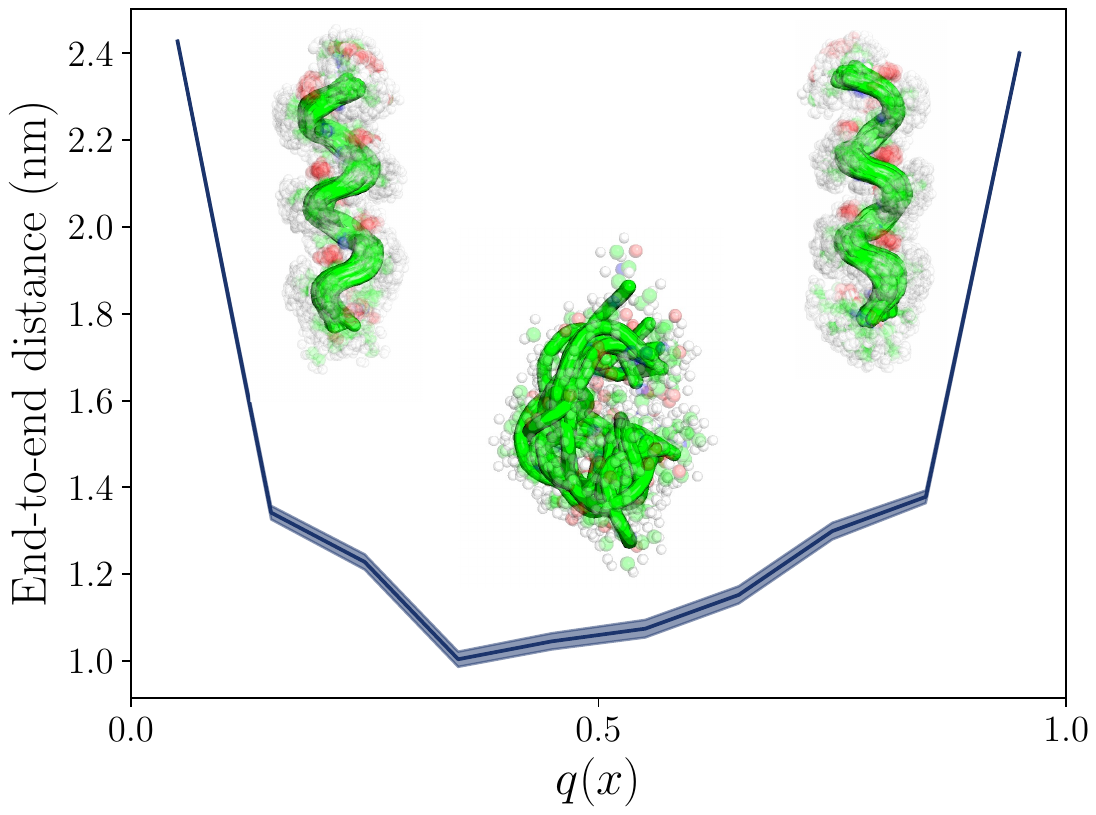}
    \caption{With only information characterizing the two metastable helical configurations of AIB9, our algorithm samples a transition state ensemble. The end-to-end distance of the transition state ensemble is shown to be much smaller than the helical configuration, suggesting that the formation of contacts between the terminal residues of AIB9 is an important step in the isomerization process.}
    \label{fig:tsensemble}
\end{figure}

Furthermore, we obtain independently useful information about the transition state through knowledge of the committor.
For example, we can sample the transition state, as shown in Fig.~\ref{fig:tsensemble}, which illustrates the non-obvious characteristic of the reaction pathway that the synthetic peptide collapses during isomerization.
This analysis illustrates a powerful use of the committor to design and test intuitive reaction coordinates with no additional simulation requirements.

\section{Conclusions}

The formulation of efficient and stable objective functions for the committor have led to numerous works seeking to learn this ideal reaction coordinate from molecular dynamics simulation data.
These algorithms build upon general purpose machine learning strategies for solving high-dimensional PDEs with a neural network ansatz~\cite{karniadakis_physics-informed_2021}. 
We have shown that combining a variational objective for the Feynman-Kac fomulation of the committor with an iterative importance sampling scheme enables accurate parameterization of this high-dimensional function.
This algorithm is agnostic to the underlying sampling distribution~\cite{li_semigroup_2022}, meaning that reweighted adaptive importance sampling can be avoided so long as the transition interfaces are well sampled.

Access to a representation of the high-dimensional committor also enables rate estimates using the Hill relation.
We show that for single barrier isomerization reactions, these estimators are computationally efficient and provide very accurate rate estimates.
Our estimator performs well in part due to the optimization of the committor using a logarithmic loss function, which we show has a statistical bias but that this bias is easily controlled. 
For a more complex reaction featuring multiple metastable intermediates, the computational cost of rate estimates remains high due to a breakdown of the assumption of Markovianity.

Because our focus in the present work is the overall sampling scheme, we intentionally used simple neural network architectures.
Of course, we believe there will be advantages in exploring the rapidly growing literature on neural network architectures appropriate for chemical systems~\cite{schutt_schnet_2018, batatia_mace_2022, gasteiger_gemnet_2022}, and it will be beneficial to assess representations that are more tailored to proteins in the future. 
Additionally, there are natural opportunities to combine the algorithms that we have developed with reaction coordinate learning~\cite{mardt_vampnets_2018, mehdi_enhanced_2024} and other enhanced sampling strategies~\cite{giardina_simulating_2011,heymann_geometric_2008}.
Furthermore, combining the approach we have taken here with algorithms more naturally suited to metastable intermediates, such as Markov state models~\cite{chodera_markov_2014}, presents a natural opportunity.

\bibliography{refs,references}

\appendix

\section{Computational Specifications and Parameter Choice}

All of our simulations were run with OpenMM 7.6.0 \cite{openmm_7}. A GitHub repository containing the code we used to run our experiments can be found at \url{https://github.com/rotskoff-group/sc\_committor\_rates}.

\subsection{Choice of lag time}

In principle, for the importance sampling phase, $\tau$ should be chosen to be on a simliar scale to the molecular relaxation time.
This prevents our committor-greedy sampling scheme from inefficiently generating correlated samples and from exploring energetically unreasonable regions of state space (as would be the case with $\tau$ too small) while still allowing for fluctuations in the direction of reaction (which would relax away with $\tau$ too large).
In practice, we have found it advantageous to avoid the small-$\tau$ regime by setting a time interval $t_\textrm{stride}$ and evolve each swarm in multiples of this interval before checking for a stopping criterion. In other words, only every $t_\textrm{stride}$ do we check the basin membership of each trajectory in the swarm, and we stop evolving the swarm if one or more trajectories has entered a basin.
Thus, $t_\textrm{stride}$ can be thought of as an effective minimum value for $\tau$.
In the case of a $t_\textrm{stride}$ equal to a single integration, we found that the stopping criterion led to swarms with $\tau$ on the order of only a few integration steps near a basin, leading to issues associated with the small-$\tau$ regime.
While choosing a larger $t_\textrm{stride}$ could theoretically bias committor estimation near the basins, our results demonstrate that such a bias has minimal effect on rate estimation. The nature of the effect of $t_\textrm{stride}$ on committor estimation is left to future work.

\subsection{Overdamped Alanine Dipeptide}

We simulated the overdamped alanine dipeptide system using OpenMM's Brownian integrator, with a time step of $0.01$ fs, a friction coefficient of $1/{\textrm{ps}}$, and hydrogen-bond restraints.
Basins A and B were defined by a circle of radius $10^\textrm{o}$ in $[\phi, \psi]$-dihedral space around the coordinates $[-150^\textrm{o}, 170^\textrm{o}]$ and $[90^\textrm{o}, -50^\textrm{o}]$, respectively. 
We ran initial equilibrium simulations in each basin and collected a boundary ensemble of 1000 configurations from each. For the sampling process, we chose swarms of size $k=100$ and maximum length $\tau = 10$ fs, with $t_\textrm{stride} = 0.1$ fs, stopping early if a swarm trajectory re-entered a basin. We ran the simulation for a total of $48$ hours on a single GPU, during which the algorithm sampled roughly $8000$ configurations; this represents $8 \times 10^5$ swarm endpoints used in the process of committor training, and about $2$ ns of total simulation time, equivalent to $2 \times 10^8$ total integration steps. For reference, the mean first passage time for the $A \longrightarrow B$ transition in this system is on the order of $10^9$ steps.

The neural network used to compute the committor was implemented in PyTorch using the architecture described in Section \ref{sec:results}. After each sampling step, the committor was trained using full-batch stochastic gradient descent using the ADAM optimizer \cite{kingma_adam_2017} with a learning rate of $10^{-4}$ for $100$ iterations. Alternate approaches that involve batching, alternative network architectures, or periodic recomputations of the target committor value $\frac{1}{k}\sum_{j=1}^k q_{\theta}(\xb_{\tau}^{\xb_i,j}))$ are possible, but we leave these to future work.

\subsection{Underdamped Alanine Dipeptide}

We simulated underdamped alanine dipeptide system using the Langevin BAOAB integrator~\cite{leimkuhler_rational_2012} from the OpenMMTools package, with a time step of $1$ fs, a friction coefficient of $1/{\textrm{ps}}$, and hydrogen-bond restraints. Basins A and B were defined identically to the overdamped case, and we collected the same number of configurations from each boundary ensemble. The swarm size was again chosen to be $k=100$; however, we chose $\tau = 100$ fs and chose the $t_\textrm{stride}$ to be an identical value; in the case of underdamped dynamics, the small $\tau$ regime is especially dangerous as the assumption of Markovianity in configuration space breaks down. Again, we terminated swarm evolution early if a swarm trajectory re-entered a basin. We also ran this simulation for a total of $48$ hours on a single GPU, sampling roughly $5800$ configurations over $55$ ns. The training regime was again identical to the overdamped case.

\subsection{Underdamped AIB9}

We also simulated the AIB9 system using the Langevin BAOAB integrator from the OpenMMTools package, with a time step of $1$ fs, a friction coefficient of $1/{\textrm{ps}}$, and hydrogen-bond restraints.
Basins A and B were defined by a circle of radius 10 in $[\phi, \psi]$-dihedral space around the coordinates $[-60, -40]$ and $[60, 40]$, respectively. 
We ran initial equilibrium simulations in each basin and collected a boundary ensemble of 1000 configurations from each. We chose swarms of size $k=100$ and $t_\textrm{stride} = 1$ ps; we did not set a fixed value for $\tau$ and instead propagated each swarm until at least one swarm trajectory re-entered a basin. We ran the simulation for a total of $96$ hours on a single GPU, during which the algorithm sampled roughly $8400$ configurations. Because of long $\tau$ in this system, a total of roughly $2.75 \, \mu$s was required to sample these configurations and their swarms. The neural network architecture used for this system mirrors that of alanine dipeptide.

\section{Comparison of rate estimates using loss with and without logarithmic correction}

To evaluate the influence of the logarithmic correction on the loss function, we compared rate estimates on a two-dimensional potential with an intermediate and two reaction channels.
We consider the potential
\begin{equation}
\label{eq:2chan}
    V(x) = \sum_{i=1}^4 A_i e^{(x-\mu_i)^2}
\end{equation}
with $A_1 = 30$, $A_2=-30$, $A_3=-50$, $A_4=-50$ and $\mu_1=(0,1/3)^T$ $\mu_1=(0,5/3)^T$, $\mu_1=(-1,0)^T$, $\mu_1=(1,0)^T$, and use overdamped dynamics with $\beta=1.$
We solved the committor function numerically exactly using the finite element method and used this exact solution to quantify the signed committor error. 
The results unambiguously demonstrate the superiority of the logarithmic objective, which we also have found to be the case in molecular transition rate estimates. 

\begin{figure}[h]
    \centering
    \includegraphics[width=0.6\linewidth]{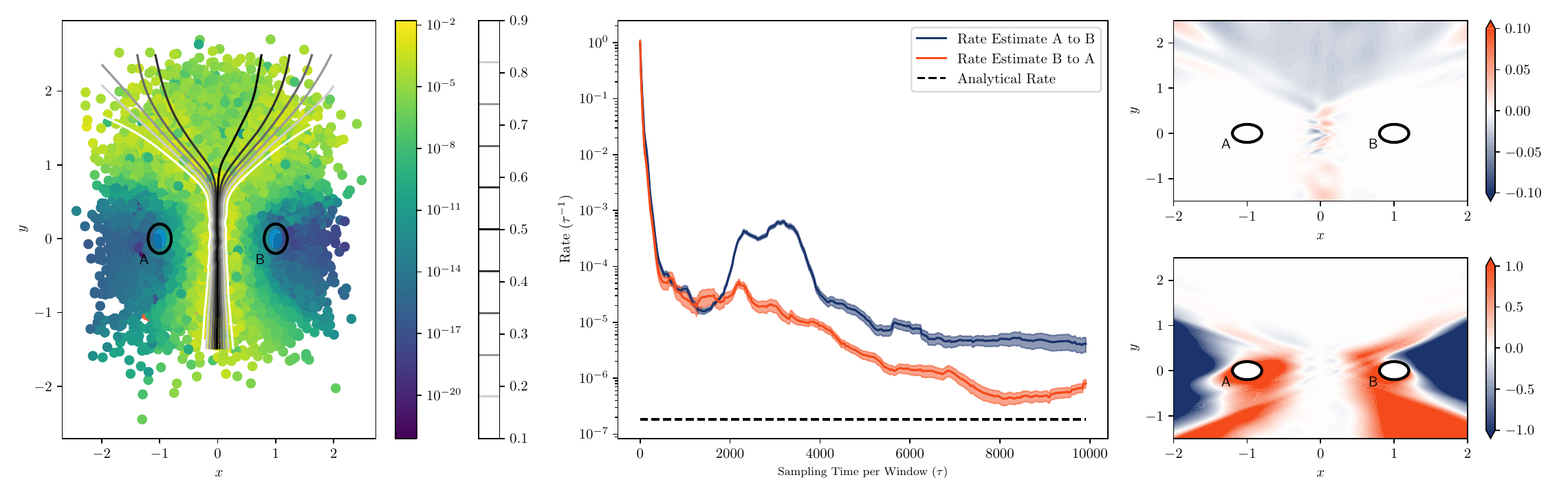}
    \includegraphics[width=0.6\linewidth]{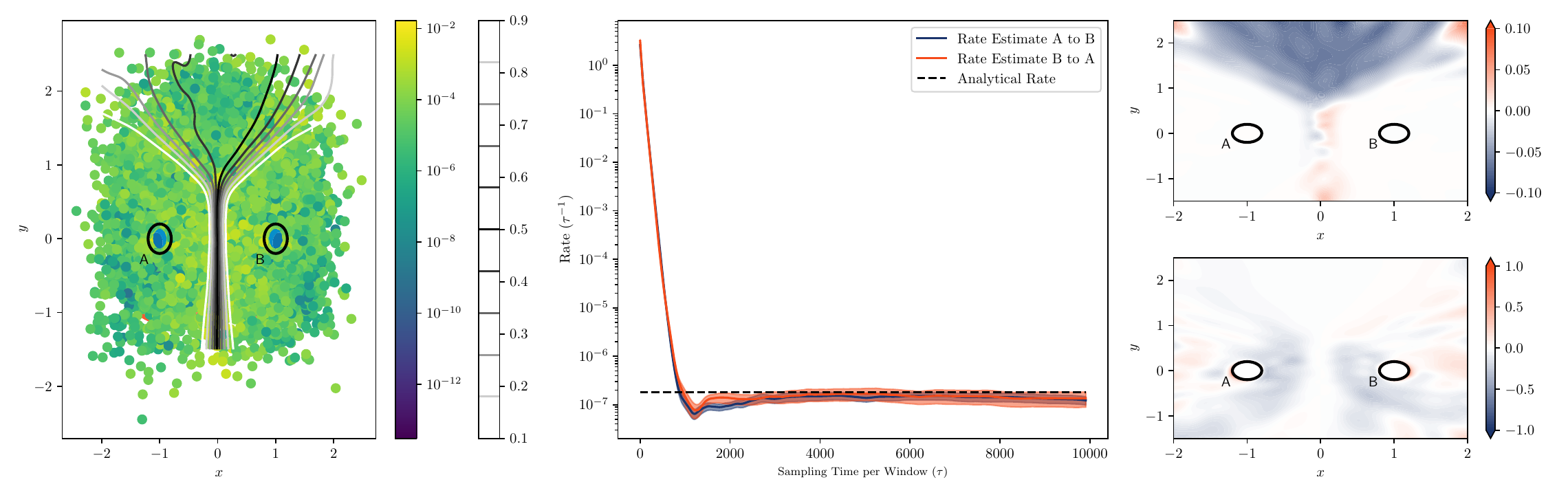}
    \caption{We directly compare optimization of the rate online calculation using the two channel potential~\eqref{eq:2chan}. Each row shows estimated committor contours alongside each configuration's relative contribution to the loss (left), the rate estimate (center), including the signed error relative to a numerically exact solution (right).
    The top row shows a committor optimized with mean squared error. The bottom row uses the logarithmic objective.}
    \label{fig:enter-label}
\end{figure}

\end{document}

%% file: preamble.tex
\usepackage[x11names, rgb]{xcolor}

\definecolor{red}{HTML}{f54b1a}
\definecolor{pink}{HTML}{d19eb1}
\definecolor{orange}{HTML}{d3772e}
\definecolor{yellow}{HTML}{ebe85d}
\definecolor{green}{HTML}{0f6852}
\definecolor{lightblue}{HTML}{01abe9}
\definecolor{darkblue}{HTML}{1b346c}
\definecolor{tan}{HTML}{e5c39e}
\definecolor{darktan}{HTML}{af9e73}
\definecolor{grey}{HTML}{c3ced0}
\definecolor{darkgrey}{HTML}{9dadc4}
\definecolor{black}{HTML}{110d1b}
\definecolor{white}{HTML}{f1f8f1}

\usepackage{hyperref}
\hypersetup{
  colorlinks = true,
  linkcolor = red,
  citecolor = darkblue,
  urlcolor = lightblue
}

\usepackage[cmintegrals,cmbraces]{newtxmath}

\usepackage{graphicx}

\usepackage[margin=1in]{geometry}

\usepackage{algorithm}
\usepackage{algpseudocode}
\algrenewcommand{\algorithmiccomment}[1]{$\vartriangleright$ #1}
\algrenewcommand{\algorithmicreturn}{\textbf{Return: }}
\algnewcommand\algorithmicinput{\textbf{Input: }}
\algnewcommand\Input{\State \algorithmicinput}

\def\qb{\boldsymbol{q}}
\def\pb{\boldsymbol{p}}

\def\xb{\boldsymbol{x}}

\def\Xb{\boldsymbol{X}}

\def\Wb{\boldsymbol{W}}

\def\d{\mathrm{d}}

\def\RR{\mathbb{R}}  
\def\EE{\mathbb{E}}

\def\<{\langle} \def\>{\rangle}